\documentclass{elsart}
\usepackage{epsfig,amstext}
\begin{document}
\begin{frontmatter}

\title{Longitudinal momentum distributions of 
$^{\textrm{16,18}}$C fragments after one-neutron 
removal from $^{\textrm{17,19}}$C\thanksref{phd}}

\date{12. August 1998}

\thanks[phd]{Part of the Doctoral Thesis of T. Baumann.}

\author[gsi]{T. Baumann},
\author[gsi]{H. Geissel},
\author[gie]{H. Lenske},
\author[goe]{K. Markenroth},
\author[gsi]{W. Schwab},
\author[goe]{M.H. Smedberg},
\author[gsi]{T. Aumann},
\author[goe]{L. Axelsson},
\author[aar]{U. Bergmann},
\author[mad]{M.J.G. Borge},
\author[gsi]{D. Cortina-Gil},
\author[mad]{L. Fraile},
\author[gsi]{M. Hellstr{\"o}m\thanksref{mh}},
\author[bra]{M. Ivanov},
\author[gsi]{N. Iwasa\thanksref{ni}},
\author[bra]{R. Janik},
\author[goe]{B. Jonson},
\author[gsi]{G. M{\"u}nzenberg},
\author[gsi]{F. Nickel},
\author[cer]{T. Nilsson},
\author[rik]{A. Ozawa},
\author[tud]{A. Richter},
\author[aar]{K. Riisager},
\author[gsi]{C. Scheidenberger},
\author[tud]{G. Schrieder},
\author[tud]{H. Simon},
\author[bra]{B. Sitar},
\author[bra]{P. Strmen},
\author[gsi]{K. S{\"u}mmerer},
\author[rik]{T. Suzuki},
\author[gsi]{M. Winkler},
\author[gii]{H. Wollnik},
\author[goe]{M.V. Zhukov}

\thanks[mh]{Present Address: Department of Physics, Lund University, 
S-22100 Lund, Sweden}
\thanks[ni]{Present Address: RIKEN, 2-1 Hirosawa, Wako, Saitama 351-01, Japan}

\address[gsi]{Gesellschaft f{\"u}r Schwerionenforschung,
              D-64291 Darmstadt, Germany}
\address[gie]{Institut f{\"u}r Theoretische Physik I,
              D-35392 Giessen, Germany}
\address[goe]{Fysiska Institutionen, Chalmers Tekniska H{\"o}gskola,
              S-41296 G{\"o}teborg, Sweden}
\address[aar]{Institut for Fysik og Astronomi, Aarhus Universitet,
              DK-8000 Aarhus C, Denmark}
\address[mad]{Instituto Estructura de la Materia, CSIC,
              E-28006 Madrid, Spain}
\address[bra]{    
              Comenius University,
              84215 Bratislava, Slovakia}
\address[cer]{EP Division, CERN, CH-1211  Gen{\`e}ve 23, Switzerland}
\address[rik]{RIKEN, 2-1 Hirosawa, Wako,
              Saitama 351-01, Japan}
\address[tud]{Institut f{\"u}r Kernphysik, Technische Universit{\"a}t,
              D-64289 Darmstadt, Germany}
\address[gii]{II. Physikalisches Institut, Universit{\"a}t Giessen,
              D-35392 Giessen, Germany}
\end{frontmatter}
\newpage
\begin{frontmatter}
\begin{abstract}
The fragment separator FRS at GSI was used as an energy-loss
spectrometer to measure the longitudinal momentum distributions of
$^{16,18}$C fragments after one-neutron removal reactions in
$^{17,19}$C impinging on a carbon target at about 910~MeV/u. The
distributions in the projectile frames are characterized by  a FWHM of
$141\pm6$~MeV/$c$ for $^{16}$C and $69\pm3$~MeV/$c$ for $^{18}$C\@.
The results are compared with experimental data obtained at lower
energies and discussed within existing theoretical models.

\end{abstract}

\end{frontmatter}
The dripline nucleus $^{19}$C has previously been studied
experimentally at MSU \cite{Baz95,Baz97} and GANIL \cite{Mar96}. The
observed momentum distributions of $^{18}$C fragments and neutrons
from breakup reactions were reported to be very narrow and interpreted
as evidence for a one-neutron halo structure in the $^{19}$C ground
state. This attracted much attention because before only one case of a
one-neutron halo nucleus has been confirmed experimentally, namely
$^{11}$Be \cite{Ann93}. The nucleus $^{11}$Be, where relative
$s$-motion dominates the ground state, has been subject to extensive
theoretical studies over the last few years
\cite{Ann94,Nun95,Esb95,Mau95,Han96} and has provided a theoretical
testing ground for one-neutron halo states.

In general, a nuclear halo state is characterized \cite{Rii92} by a
low separation energy and low angular momentum for the valence
nucleon(s). The $^{17,19}$C isotopes have small one-neutron separation
energies compared to those of the corresponding core. As an example,
$S_{n}(^{19}\textrm{C})=242\pm95$~keV\footnote{This value is based on
the four existing mass determinations \cite{Mar96}, while the latest
{\sc Nubase} evaluation \cite{Aud97} gives the value
$S_{n}(^{19}\textrm{C})=160\pm110$~keV, based only on the two most
recent measurements.} \cite{Baz95} which is an order of magnitude
smaller than that of the core,
$S_{n}(^{18}\textrm{C})=4180\pm30$~keV. Although the experimental
determination of the $^{19}$C mass is not very accurate, the
neutron separation energy appears to be even smaller than the one for
$^{11}$Be ($S_{n}(^{11}\textrm{Be})=503\pm6$~keV \cite{Aud97}), which
is a strong motivation for investigating the nuclear structure
of $^{19}$C\@.

\begin{figure}
\centering{\epsfig{file=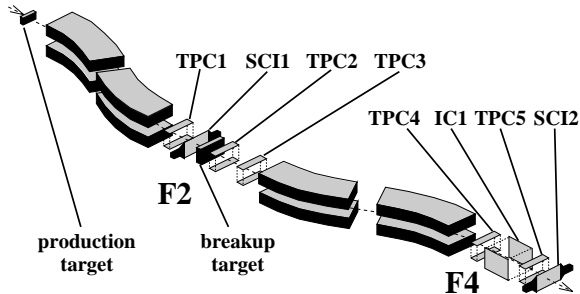,width=7.6cm}}
\caption{
Experimental setup for longitudinal momentum measurements with the FRS
operated in energy-loss mode. Complete tracking of the fragments at
the central focal plane (F2) and the final focus (F4) was achieved by
using five time projection chambers (TPC1--TPC5). Plastic
scintillators (SCI1 \& SCI2) provided time-of-flight and $\Delta E$
information. Additionally, an ionization chamber (IC1) was used for
the identification of the proton number.}
\label{fig1}
\end{figure}

In the present study, we have carried out measurements of the
longitudinal momentum distribution of $^{16,18}$C fragments after
breakup of $^{17,19}$C\@. The heavy-ion synchrotron SIS at GSI
delivered an $^{40}$Ar primary beam at 1~GeV/u with an intensity of
about $8 \cdot 10^{9}$ particles per spill. For the production of the
secondary beam, a $^{9}$Be target of 6.33~g/cm$^2$ thickness was
placed at the entrance of the magnetic spectrometer FRS \cite{Gei92}.
Figure~\ref{fig1} shows the experimental arrangement at the FRS, which
was operated as an energy-loss spectrometer. In this ion-optical mode,
the momentum distribution after reactions in the breakup target can be
measured independently of the relatively large momentum spread of the
incident secondary beam. The breakup target, 4.45~g/cm$^{2}$ carbon,
was situated at the central focal plane (F2), where the dispersion was
7.1~m. The final focus (F4) was achromatic with respect to the
production target, but had a dispersion of 5.9~m with respect to the
breakup target.

For a complete tracking of the incident particles and the breakup
fragments at the central focal plane, three gas-filled time projection
chambers (TPC1--TPC3) were employed, one placed in front of the
breakup target and two placed behind it. The TPC is a two-dimensional
position sensitive detector with a homogeneous matter distribution and
a resolution better than 0.5~mm in $x$- and $y$-coordinates
\cite{Bau96}. The longitudinal momentum induced at the breakup target
was determined from the positions in TPC4 and TPC5 at the final focal
plane.

\begin{figure}
\centering{\epsfig{file=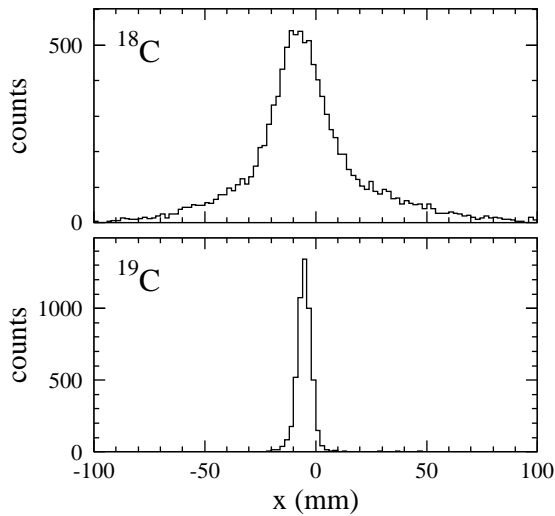,width=7.6 cm}}
\caption{
Measured position distributions of $^{18,19}$C nuclei at F4. The
distribution of the $^{18}$C nuclei results from the one-neutron
removal reaction of $^{19}$C, whereas the $^{19}$C distribution
corresponds to a different $B\rho$ setting of the FRS selecting the
fragments which penetrated the breakup target without nuclear
reactions.
}
\label{fig2}
\end{figure}

For the measurements of the breakup reactions, the magnetic fields of
the first two dipole stages of the FRS, including quadrupole and
hexapole magnets, were set to select the beams of $^{17}$C or
$^{19}$C\@. A scintillation detector (SCI1) in front of the breakup
target was used to identify the proton number $Z$ of the incoming
particles. The magnets behind the breakup target were set to a
magnetic rigidity ($B\rho$) corresponding to fragments arising from a
one-neutron removal of the selected projectile. Particles arriving at
the final focus (F4) were identified by measuring the time-of-flight
between the scintillators SCI1 and SCI2, by determining the magnetic
rigidity from the position measurement, and by a coincident
energy-deposition measurement in an ionisation chamber (IC1). The
different isotopes were well separated in an $A/Z$ versus $Z$ plot.
The unambiguous identification at F4, the $Z$-identification in
front of the breakup target, and the individual $B\rho$-settings
ensured that only reaction products from a one-neutron removal
contributed to the measured momentum distributions. The
$B\rho$-measurement was calibrated using the primary beam at different
energies. The achromatism and the dispersions of the ion-optical
system were experimentally deduced by transmitting the $^{17}$C and
the $^{19}$C beam to the final focal plane (Fig.~\ref{fig2}, bottom
panel).

The momentum distribution of the core fragments is directly deduced
from the position measurements with TPC4 and TPC5 using the
experimentally determined dispersion at the final focus. Ion-optical
aberrations, atomic energy straggling, and angular straggling in the
relatively thick breakup target limit the resolution that can be
achieved in this measurement. These contributions, which broaden the
measured widths by about 3\%, were quantified by measuring the
position distributions of $^{17}$C and $^{19}$C nuclei that penetrated
the breakup target without nuclear reactions. In Fig.~\ref{fig2}, the
measured position distribution of $^{19}$C at F4 is presented together
with the corresponding distribution of the $^{18}$C fragments.

The breakup of the projectile nuclei is statistically distributed
along their path inside the target. This causes an additional energy
straggling due to the difference in energy loss between the nuclei
before and after one-neutron removal. For this, we computed a negligible 
contribution for the momentum of $\Delta p/p=0.5\cdot10^{-3}$
by applying the ion-optical ray tracing code {\sc Mocadi} \cite{Iwa97}.

\begin{figure}
\centering{\epsfig{file=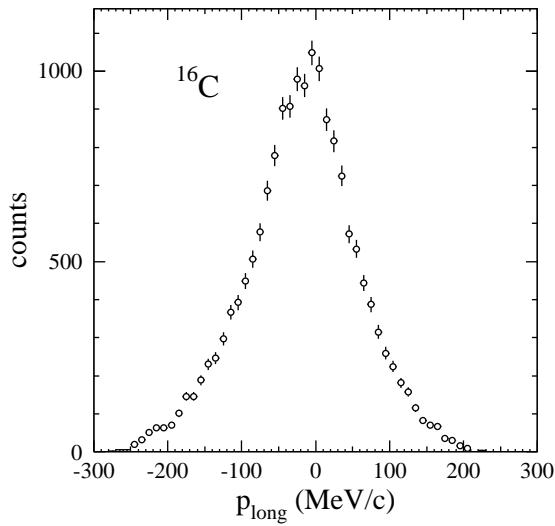,width=7.6 cm}}
\caption{
Momentum spectrum of $^{16}$C after one-neutron removal from
$^{17}$C\@. The FWHM, corrected for the experimental resolution, is
$141\pm6$~MeV/$c$.
}
\label{fig3}
\end{figure}

The longitudinal momentum distribution of $^{16}$C fragments from the
breakup of $^{17}$C is shown in Fig.~\ref{fig3}.
         This momentum distribution, which is transformed into the
         $^{17}$C projectile frame, exhibits neither a pure Lorentzian
         nor a Gaussian distribution. Using a combination of two
         Gaussian distributions, we could reproduce the shape and
         determined a FWHM of $145\pm6$~MeV/$c$.
After correction for the experimental resolution mentioned above, this
gives a FWHM of $141\pm6$~MeV/$c$. This value agrees very well with a
recent experiment at 84~MeV/u beam energy, where a FWHM of 
$145\pm5$~MeV/$c$ was obtained \cite{Baz97}.
         There also exists a published result from an earlier
         measurement of $^{17}$C at low energies \cite{Baz95}, but
         this value suffers from a poor identification as stated in
         Ref.~\cite{Baz97} and should not be used for this comparison.

\begin{figure}
\centering{\epsfig{file=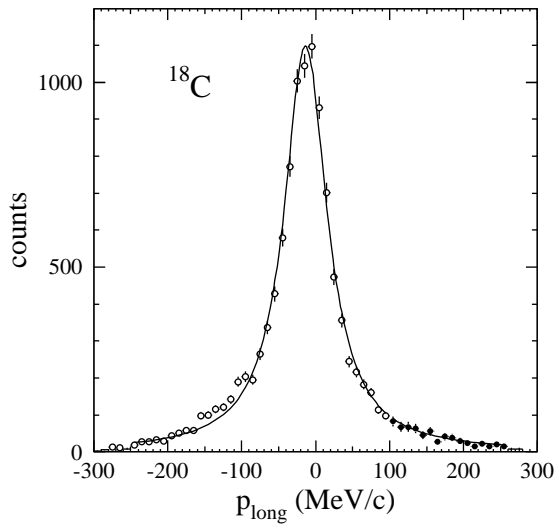,width=7.6 cm}}
\caption{
Momentum spectrum of $^{18}$C after one-neutron removal from
$^{19}$C\@. The distribution, transformed to the reference frame of
the $^{19}$C fragments, consists of a combination of two spectra
recorded at two $B\rho$ settings of the fields to include also the
tails of the distribution. The result of a Lorentzian fit with a FWHM
of $71\pm3$~MeV/$c$ is shown as a solid line. With the correction for
experimental resolution taken into account, the resulting FWHM is 
$69\pm3$~MeV/$c$.
}
\label{fig4}
\end{figure}

Figure~\ref{fig4} presents the momentum distribution of $^{18}$C,
transformed into the $^{19}$C projectile frame. In order to cover a
larger range of momenta, this distribution was recorded at two
different $B\rho$-settings for the magnetic stages behind the breakup
target. The two spectra that form the combined distribution were
normalized using the total particle count rate at F2.

The shape of the momentum distribution of $^{18}$C fragments is well
reproduced by a Lorentzian between $-250$~MeV/$c$ and $+250$~MeV/$c$.
The FWHM of this distribution is  $71\pm3$~MeV/$c$, yielding
$69\pm3$~MeV/$c$ with the experimental resolution taken into account.
This width is about 3 times smaller than the one obtained from the
statistical model of Goldhaber \cite{Gol74} with the refined
parameterization of Morrissey \cite{Mor89}, which in general gives a
good description of experimental results for tightly bound nucleons.

The measured width for the $^{19}$C breakup agrees with the neutron
momentum width of  $64\pm17$~MeV/$c$ (FWHM) obtained in an experiment
at 30~MeV/u \cite{Mar96}.  However, it is significantly larger than
the result of the investigation of $^{19}$C at an energy of 77~MeV/u,
where a FWHM of $42\pm4$~MeV/$c$ (c.m.\ frame) was reported.
         This value is based on the first measurement of the nuclear
         breakup of $^{19}$C (see Ref.~\cite{Baz95}), but the data
         have been revised in a later publication \cite{Baz97}, to
         which we refer here.

The main differences between Refs.~\cite{Baz95,Baz97} and our
measurement are the projectile energies (77~MeV/u and 914~MeV/u) and
the target materials (beryllium and  carbon). In earlier measurements
of longitudinal momentum distributions for the halo nuclei $^{11}$Be
and $^{11}$Li, no significant dependence of the width on beam energy
or target material has been observed.
         The effects of a limited spectrometer acceptance were taken
         into account for the results of Ref.~\cite{Baz97}. In the
         case of our measurement, these effects are negligible due to
         the forward focussing at relativistic beam energies
         \cite{Gei97}. Nevertheless, this was carefully checked in our
         analysis using the complete particle tracking at the breakup
         target and at the final focal plane. From the agreement of
         the results for $^{17}$C between Ref.~\cite{Baz97} and our
         measurement, we conclude that systematic experimental effects
         can be ruled out as a cause of the difference observed for
         $^{19}$C\@. One should keep in mind, however, that the
         statistics of Ref.~\cite{Baz97} for $^{19}$C are very low. A
         measurement at low energies with comparable statistics would
         be desirable in order to confirm or disprove this difference. 
We note for later use that the presently measured width for $^{19}$C is
considerably larger than the 43--50~MeV/$c$ \cite{Kel95,Gei97} found
for $^{11}$Be in spite of the one-neutron separation energy being a
factor of about two lower in the case of $^{19}$C\@.

It is not the aim of the present paper to provide a new attempt for a
theoretical description of the structure of $^{17,19}$C since there
are several recent works that have dealt with this problem. Instead,
we shall give a brief outline of the present status and comment on
what we can conclude at this stage with respect to our experimental
findings.

\begin{figure}
\centering{\epsfig{file=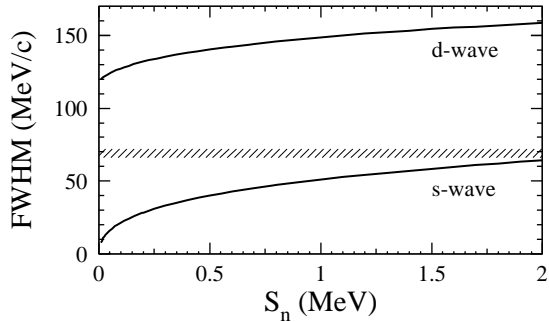,width=7.6 cm}}
\caption{
Calculated momentum widths as a function of the neutron binding energy
$S_{n}$ for $^{18}$C fragments after breakup of $^{19}$C\@. The
calculated values are for pure $s$- and $d$-orbits and the points
cover the range of separation energies representing a coupling of the
$s$- and $d$-state to the $0^+$ ground state or the $2^+_1$ state at
1.62~MeV. Note that the effective neutron separation energy of a
neutron coupled to the $2^+_1$ state is 1.86~MeV. Hankel wave
functions with a cylindrical cutoff parameter of 5.45~fm were used
in this calculation. The hatched region marks the experimental width
obtained in this work.
}
\label{fig5}
\end{figure}

A common feature of $^{17,19}$C is a loosely bound neutron moving
outside a core which itself is already far off stability. The
existence of low-lying $2^+$ states in $^{16,18}$C at 1.77 and 1.62~MeV
\cite{Til93,Til95}, respectively, indicates that core polarization
may play an important role in these nuclides. The ground state wave
function of $^{17,19}$C can then be expected to have $1s_{1/2}$ and
$0d_{3/2,5/2}$ neutrons coupled to the $0^+$ ground state and the
$2^+_1$ state as the dominating configurations. To illustrate the
contributions to the experimental widths from the different components
of the wave function, we show in Fig.~\ref{fig5} a calculation of the
momentum widths after breakup of $^{19}$C for pure $s$- and $d$-orbits
as a function of the neutron separation energy. In this calculation, we
used Hankel radial wave functions, which are the exact solutions
of the Schr{\"o}dinger equation outside the range of the potential,
and introduced a lower cylindrical cutoff when transforming it to the
corresponding momentum coordinates. This cutoff, estimated as the sum
of the core and target radii \cite{Han96}, ensures that the
one-neutron removal process takes place in the outer region of the
wave function, keeping the core intact. The measured longitudinal
momentum distribution of the $^{18}$C fragment favors the
$J^{\pi}=3/2^{+}$ or $5/2^{+}$ scenario where the main part of the
wave function ($\sim65$\%) contains relative $s$-motion between the
halo neutron and the $2_{1}^+$ excited state of $^{18}$C\@. The
experimental data for one-neutron removal from $^{17}$C can be
reproduced by an almost pure $d$-wave neutron ($S_{n}=0.73$~MeV)
orbiting around the $0^+$ ground state of $^{16}$C\@. Inclusion of the
first excited $2^+_1$ state at 1.77~MeV of $^{16}$C further improves
the agreement with the experimental data.

Bazin {\it et al.} also used Hankel wave functions to fit their data
\cite{Baz95,Baz97}. They used shell model calculations to determine
the parentage of the ground state to single particle wave functions.
The best agreement with the experimental data for $^{17}$C was found
when a spin parity of $3/2^+$ from $s$- and $d$-wave neutrons coupled
to the $2^+_1$ state in $^{16}$C was assumed. In the $^{19}$C case,
they used the WBT interaction predicting the spin parity of the ground
state to be $5/2^+$ with the dominating configuration
$^{18}\textrm{C}(2^+_{1})\otimes1s_{1/2}$. With these assumptions, they
claimed to get a good fit to the width, while from our calculation
(Fig.~\ref{fig5}) we would expect a value closer to 60 MeV/$c$.

Ridikas {\it et al.} \cite{Rid97,Rid97a} investigated $^{17,19}$C
in a neutron-plus-core coupling model with a deformed Wood-Saxon
potential for the neutron--core interaction. Their analysis excluded a
$J^{\pi}=1/2^+$ for $^{17}$C, while in the $^{19}$C case, they concluded
that the main contribution to the ground state comes from relative
$s$-motion between the halo neutron and the core, with the $^{18}$C
core either in its ground state or the $2^+_1$ state. The calculations
in Refs.~\cite{Rid97,Rid97a} were compared to the experimental FWHM of
$\approx41$~MeV/$c$ \cite{Baz95}, but there were difficulties to
reproduce such a narrow width \cite{Rid97a}. A better agreement with
the value given in this paper is obtained if $J^{\pi}=3/2^+, 5/2^+$ is
assumed, and if the wave function has an appreciable amount of
$s$-motion coupled to the $2^+$ state of $^{18}$C\@.

Describing the coupling of single particle configurations to excitations
of the core nucleus by dynamical core polarization (DCP) leads to a more
detailed approach. As outlined in Ref.~\cite{Len97} and discussed for
stable nuclei in Refs.~\cite{Eck89,Eck90,Bre88,Neu90}, the core--particle
interactions are obtained microscopically from RPA calculations. This
leads to non-static and non-local self-energies thus extending the
nuclear mean-field description beyond the static Hartree-Fock approach.
Dynamical core polarization is well understood for stable nuclei, where
it is found to describe rather accurately single particle strength
distributions of odd mass nuclei as, e.\,g., in
Refs.~\cite{Eck89,Eck90,Bre88,Neu90}.

In the DCP approach, a $1/2^+$ ground state for $^{17}$C is found
which, however, is almost degenerate with a nearby $5/2^+$ state at
$E_{\textrm{x}}=70$~keV. For $^{19}$C, the calculations predict 
a $1/2^+$ ground state obtained at a separation energy
$S_{n}=183$~keV\cite{Len98}. 
The first $5/2^+$ state is found about 300~keV above
the ground state, and is located just beyond the continuum threshold.
The $^{18}\textrm{C}(0^+)\otimes1s_{1/2}$ leading particle
configuration accounts for only 40\% of the wave function. The
$^{18}\textrm{C}(2^+_{1})\otimes0d_{5/2}$ core excited configuration
accounts for the major part of the $^{19}$C ground state. These
results indicate that the binding of $^{19}$C is probably dominated by
dynamical particle--core interactions rather than static mean-field
dynamics. As a consequence, shell structures are dissolved and the
last neutron is distributed over the single particle orbitals in the
($1s,0d$)-valence shell coupled to core excited configurations.

It is clear that the proximity of the $s$- and $d$-states makes it
difficult to predict the ground state configurations based only on
measured momentum distributions. The half-life predictions for
$^{19}$C are not sensitive to the ground state spin \cite{Baz95}, and
such experimental information is therefore non-conclusive. The various
theoretical approaches---although being very different in
detail---lead to the common conclusion that the first excited $2^+_1$
state of the core plays an important role in both $^{17}$C and
$^{19}$C\@. If so, the spatial extension of the halo neutron would be
less than indicated by the ground state separation energy alone. The
above comparison of the widths obtained for $^{19}$C and $^{11}$Be
indeed shows that the $^{19}$C ground state is a less developed
one-neutron halo state.

For future experimental work, one of the challenges will be a
determination of the contribution from the first excited core state,
either by the observation of $\gamma$ rays in coincidence with the
charged fragments, or by high-resolution mass spectroscopy. Another
task is to verify the experimental indication of an energy
dependence of the widths, and, if this trend remains, to give a
theoretical description of it.

\section*{Acknowledgements}
This work was supported by the German Federal Minister for Education
and Research (BMBF) under Contracts 06~DA~820, 06~OF~474, and
06~MZ~476 and by GSI via Hochschulzusammenarbeitsvereinbarungen under
Contracts DARIK, OF~ELK, and MZ~KRK\@. It was partly supported by the
Polish Committee of Scientific Research under Contract
PB2/P03B/113/09, EC under Contract ERBCHGE-CT92-0003, CICYT under
Contract AEN92-0788-C02-02 (MJGB), and by Deutsche
Forschungsgemeinschaft (DFG) under Contract 436~RUS~130/127/1. One of
us (B.J.) acknowledges the support through an Alexander von Humboldt
Research Award.

\end{document}